\documentstyle[epsf,12pt]{article}

\newtheorem{theorem}{\bf Theorem}
\newtheorem{remark}{\sl Remark}

\def\aligned#1{\begin{array}{#1}}
\def\endaligned{\end{array}}

\def\matrix#1{\begin{array}{#1}}
\def\endmatrix{\end{array}}
\def\pmatrix#1{\left(\begin{array}{#1}}
\def\endpmatrix{\end{array}\right)}

\def\D{\displaystyle}
\def\tr{\mathop{\hbox{\,\rm tr}}\nolimits}
\def\re{\mathop{\hbox{\,\rm Re}}\nolimits}
\def\im{\mathop{\hbox{\,\rm Im}}\nolimits}
\def\hc{\hbox{\bf C}}
\def\hr{\hbox{\bf R}}
\def\diag{\mathop{\hbox{\,\rm diag}}\nolimits}

\def\d{\hbox{\rm d}}

\newif\ifinexp \inexpfalse
\def\I{\ifinexp \hbox{\hskip0.8pt\eightrm i} \else \hbox{\hskip1pt\rm i} \fi}

\newcount\egcount \egcount=1
\def\example{\vskip4pt
   \noindent{\bf Example \the\egcount:}\par \advance\egcount by 1}
\def\endexample{\vskip4pt}

\def\adS{2+1 dimensional anti-de Sitter space-time}
\def\YMH{Yang-Mills-Higgs}

\def\figuretype{EPS}

\begin{document}

\title{Solutions of the \YMH{} equations\\ in \adS}

\author{\small Zixiang Zhou\\
\small Institute of Mathematics, Fudan University,
Shanghai 200433, China\\
\small Email: zxzhou@guomai.sh.cn}

\date{}

\maketitle

\begin{abstract}
The solutions of the Bogomolny equation in \adS{} are obtained by
using Darboux transformations with both constant spectral
parameters and variable ``spectral parameters''. These solutions
give the \YMH{} fields in \adS{}. Some examples in SU(2) case are
considered and qualitative asymptotic behaviors of the solutions
as $t\to\infty$ are discussed in detail.
\end{abstract}

\section{Introduction}\label{sec1}

The \YMH{} fields satisfying the Bolgomolny equations in $\hr^3$ and
$\hr^{2,1}$ were widely investigated and the equations are known
to be integrable. On the other hand, the \YMH{} fields satisfying the
Bolgomolny equations in some curved spaces such as the hyperbolic
space $H^3$, the \adS{} are also integrable
\cite{bib:Atiyah,bib:Hitchinbook,bib:Wardnew}. In the present
paper, we consider the solutions of the Bolgomolny equation in
the \adS, the Lax pair of which has been known and the soliton
solutions (with constant spectral parameters) were also obtained
\cite{bib:Wardnew}.

With the Darboux transformation method, we obtain exact
multi-soliton solutions. Moreover, the ``spectral
parameters'' in the construction of Darboux transformation can
depend on the space-time variables as in some other problems with
dimensions $\ge 3$, like the self-dual Yang-Mills equation,
modified principal chiral field, the Bolgomolny equation in
$\hr^3$ etc. \cite{bib:Gu,bib:Ma,bib:Ward1,bib:Zhou,bib:Gumono}.
This kind of equations are also called breaking soliton
equations, since the spectral parameter may not be constant and
satisfies an equation whose solution can blow up at finite time
\cite{bib:Li,bib:Lou}.

In Sec.~\ref{sec2} and Sec.~\ref{sec3} of the present
paper, the Darboux transformations for $GL(N,\hc)$ and $U(N)$
cases are discussed. As a special case, the general construction
of soliton solutions is given in Sec.~\ref{sec4}. Then, in
Sec.~\ref{sec5}, some examples of single solitons and
multi-solitons are considered, with both constant spectral
parameters and variable ``spectral parameters''. Their
qualitative asymptotic behaviors as $t\to\infty$ are discussed in
detail. When the spectral parameters are constants, we find
solutions globally defined on the whole \adS. When the ``spectral
parameters'' are not constants, the solutions derived here are
only locally defined in \adS.

Our problem is as follows.

Let $M$ be a three dimensional Lorentz manifold with metric $g$.
$A_\mu$ is a gauge potential and $\Phi$ is a (scalar) Higgs
field, both of which are valued in the Lie algebra of a Lie group
$G$. Hereafter, we always suppose $G$ is a matrix Lie group and
the matrices in $G$ are of order $N$.

The \adS{} is the universal covering space of the
hyperboloid
\begin{equation}
   U^2+V^2-X^2-Y^2=1
   \label{eq:hyperboloid}
\end{equation}
in $\hr^{2,2}$ with the metric
\begin{equation}
   ds^2=-dU^2-dV^2+dX^2+dY^2.
\end{equation}
Define
\begin{equation}
   r=\frac 1{U+X},\quad
   x=\frac Y{U+X},\quad
   t=-\frac V{U+X},
   \label{eq:xrt}
\end{equation}
then a part of the \adS{} with $U+X>0$ is represented by the
Poincar\'e coordinates $(r,x,t)$ with $r>0$ and the metric is
\begin{equation}
   ds^2=r^{-2}(-dt^2+dr^2+dx^2)=r^{-2}(dr^2+du\,dv)
   \label{eq:metric}
\end{equation}
where $u=x+t$, $v=x-t$.

The \YMH{} field in \adS{} satisfies the Bogomolny equation
\cite{bib:Atiyah,bib:Hitchin1}
\begin{equation}
   D\Phi=*F,
   \label{eq:YMeq}
\end{equation}
or, written in terms of the components,
\begin{equation}
   D_\mu\Phi=\frac 1{2\sqrt{|g|}}g_{\mu\nu}
   \epsilon^{\nu\alpha\beta}F_{\alpha\beta}
   \label{eq:YMeqcompo}
\end{equation}
where the action of the covariant derivative
$D_\mu=\partial_\mu+A_\mu$ on $\Phi$ is
$D_\mu\Phi=\partial_\mu\Phi+[A_\mu,\Phi]$,
$\partial_\mu=\partial/\partial x^\mu$. $\{F_{\mu\nu}\}$ is the
curvature corresponding to $\{A_\mu\}$, $F_{\mu\nu}=[D_\mu,D_\nu]$.

With the Poincar\'e coordinates (\ref{eq:xrt}),
(\ref{eq:YMeqcompo}) becomes
\begin{equation}
   D_u\Phi=rF_{ur},\quad
   D_v\Phi=-rF_{vr},\quad
   D_r\Phi=-2rF_{uv}.
   \label{eq:EQ}
\end{equation}

It was proposed in \cite{bib:Wardnew} that this system of
nonlinear partial differential equations had a Lax pair
\begin{equation} \aligned{l}
   (rD_r+\Phi-2(\zeta-u)D_u)\psi=0,\\
   \D\left(2D_v+\frac{\zeta-u}rD_r-\frac{\zeta-u}{r^2}\Phi\right)\psi=0
\endaligned \label{eq:lp} \end{equation}
where $D_\mu\psi=\partial_\mu\psi+A_\mu\psi$ and $\zeta$ is a
complex spectral parameter. That is, (\ref{eq:EQ}) is the
integrability condition of the over-determined system
(\ref{eq:lp}).

\section{Darboux transformation in $GL(N,\hc)$ case}\label{sec2}

In this section, we consider the case $G=GL(N,\hc)$. This is the
simplest case because no reduction should be considered. Let
\begin{equation}
   \widetilde\psi=(\zeta-u)R\psi-T\psi
   \label{eq:dm}
\end{equation}
where $R(u,v,r)$ and $T(u,v,r)$ are $N\times N$ matrices and $R$
is invertible, then the transformation $\psi\to\widetilde\psi$ is
called a Darboux transformation (of degree one) if there are
$(\widetilde A_\mu, \widetilde\Phi)$ such that
\begin{equation} \aligned{l}
   (r\widetilde D_r+\widetilde\Phi
    -2(\zeta-u)\widetilde D_u)\widetilde\psi=0,\\
   \D\left(2\widetilde D_v+\frac{\zeta-u}r\widetilde D_r
    -\frac{\zeta-u}{r^2}\widetilde\Phi\right)\widetilde\psi=0
\endaligned \label{eq:lpp} \end{equation}
hold. Here $\widetilde D_\mu=\partial_\mu+\widetilde A_\mu$.

We should first determine $R$ and $T$ so that (\ref{eq:dm}) is a
Darboux transformation. For given $(A,\Phi)$, $(\widetilde
A,\widetilde\Phi)$ and arbitrary matrix function $Q$, let
\begin{equation} \aligned{l}
   \Delta_\mu Q=\partial_\mu Q+\widetilde A_\mu Q-QA_\mu,\\
   \delta Q=\widetilde\Phi Q-Q\Phi.
\endaligned \end{equation}

Expressed in $\psi$, both equations of (\ref{eq:lpp}) are
polynomials of $\zeta$ of degree two. The coefficients of the
second, first and zero-th order of $\zeta$ in the two equations
of (\ref{eq:lpp}) lead to
\begin{equation} \aligned{l}
   \Delta_uR=0,\quad r\Delta_rR+2\Delta_uT+\delta R+2R=0,\\
   r\Delta_rT+\delta T=0,
\endaligned \end{equation}
and
\begin{equation} \aligned{l}
   \D\Delta_rR-\frac 1r\delta R=0,\quad
    2\Delta_vR-\frac 1r\Delta_rT+\frac 1{r^2}\delta T=0,\\
   \Delta_vT=0.
\endaligned \end{equation}
These two systems are equivalent to
\begin{eqnarray}
   &&\Delta_uR=0, \label{eq:dteq1}\\
   &&\Delta_vT=0, \label{eq:dteq2}\\
   &&\Delta_vR=\frac 1r\Delta_rT, \label{eq:dteq3}\\
   &&\Delta_rR+\frac 1r\Delta_uT+\frac 1rR=0, \label{eq:dteq4}\\
   &&\Delta_rR-\frac 1r\delta R=0, \label{eq:dteq5}\\
   &&\Delta_rT+\frac 1r\delta T=0. \label{eq:dteq6}
\end{eqnarray}

$\widetilde A_u$ and $\widetilde A_v$ are solved from
(\ref{eq:dteq1}) and (\ref{eq:dteq2}) as
\begin{eqnarray}
   &&\widetilde A_u=RA_uR^{-1}-(\partial_uR)R^{-1}, \label{eq:Aueq}\\
   &&\widetilde A_v=TA_vT^{-1}-(\partial_vT)T^{-1},
\end{eqnarray}
while (\ref{eq:dteq5}) and (\ref{eq:dteq6}) lead to
\begin{eqnarray}
   &&\widetilde A_r=\frac 12(TA_r-\partial_rT)T^{-1}
    +\frac 12(RA_r-\partial_rR)R^{-1}
    +\frac 1{2r}(T\Phi T^{-1}-R\Phi R^{-1}),\\
   &&\widetilde\Phi=\frac r2(TA_r-\partial_rT)T^{-1}
    -\frac r2(RA_r-\partial_rR)R^{-1}
    +\frac 12(T\Phi T^{-1}+R\Phi R^{-1}).\label{eq:Phieq}
\end{eqnarray}

Now let $Z$ be an $N\times N$ matrix function of $(u,v,r)$, $H$
be a solution of
\begin{equation} \aligned{l}
   r((\partial_rH)H^{-1}+A_r)+\Phi-2((\partial_uH)H^{-1}+A_u)S=0,\\
   \D 2((\partial_vH)H^{-1}+A_v)+\frac 1r((\partial_rH)H^{-1}+A_r)S
    -\frac 1{r^2}\Phi S=0,\\
\endaligned \label{eq:Heq} \end{equation}
where $S=HZH^{-1}-u$, then $S$ satisfies
\begin{equation} \aligned{rl}
   \partial_rS=&\D H(\partial_rZ-\frac 2r(\partial_uZ)(Z-u))H^{-1}+\frac 2rS\\
   &\D+\frac 2r(\partial_uS)S-[A_r,S]+\frac 2r[A_u,S]S-\frac 1r[\Phi,S],\\
   \partial_vS=&\D H(\partial_vZ+\frac 1{2r}(\partial_rZ)(Z-u))H^{-1}\\
   &\D-\frac 1{2r}(\partial_rS)S-[A_v,S]-\frac 1{2r}[A_r,S]S
    +\frac 1{2r^2}[\Phi,S]S.\\
\endaligned \label{eq:Seq} \end{equation}

\begin{remark}
If $Z$ is diagonal and $Z=\diag(\zeta_1,\cdots,\zeta_N)$, then
$H=(h_1,\cdots,h_N)$ where $h_i$ is a column solution of the
Lax pair (\ref{eq:lp}) with $\zeta=\zeta_i(u,v,r)$.
\end{remark}

If $T=RS$, then (\ref{eq:dteq3}) and (\ref{eq:dteq4}) hold if and
only if
\begin{equation} \aligned{l}
   \D\partial_rZ-\frac 2r(\partial_uZ)(Z-u)=0,\\
   \D\partial_vZ+\frac 1{2r}(\partial_rZ)(Z-u)=0.
\endaligned \label{eq:Zeq} \end{equation}
Therefore, we have

\begin{theorem}\label{thm1}
Suppose $R(u,v,r)$ is an arbitrary invertible matrix function. If
$Z(u,v,r)$ is an $N\times N$ matrix solution of (\ref{eq:Zeq}) and
$H$ is a solution of (\ref{eq:Heq}) with $S=HZH^{-1}-u$, then
$\psi\mapsto \widetilde\psi=(\zeta-u)R\psi-RS\psi$ is a Darboux
transformation for (\ref{eq:lp}).
\end{theorem}

If $Z=\diag(\zeta_1,\cdots,\zeta_N)$, then each $\zeta_i$ is a
solution of
\begin{equation} \aligned{l}
   \D\partial_r\zeta-\frac 2r(\zeta-u)\partial_u\zeta=0,\\
   \D\partial_v\zeta+\frac 1{2r}(\zeta-u)\partial_r\zeta=0.
\endaligned \label{eq:zetaeq} \end{equation}
Apart from the constant solution, the general non-constant
solution of (\ref{eq:zetaeq}) is given implicitly by
\begin{equation}
   v-\frac{r^2}{\zeta-u}=C(\zeta)
   \label{eq:exprzeta2}
\end{equation}
where $C$ is an arbitrary holomorphic function. We still call
$\zeta_i$ as a ``spectral parameter''. However, here the
``spectral parameters'' $\zeta_i$'s in $Z$ can be either constant
or variable. In the latter case, they are given by
(\ref{eq:exprzeta2}).

\begin{remark}
Note that the spectral parameter $\zeta$ in the Lax pair is still
a complex constant. Only $\zeta_i$'s in $Z$ can depend on $(u,v,r)$.
\end{remark}

According to Theorem~\ref{thm1}, we can get exact solution of
(\ref{eq:EQ}) from a known solution of (\ref{eq:EQ}) and the
corresponding solution of the linear system (\ref{eq:Heq}). When
$Z$ is diagonal, solving (\ref{eq:Heq}) is equivalent to solving
(\ref{eq:lp}).

Theorem~\ref{thm1} gives the construction of Darboux
transformations of degree one. The Darboux transformations of
higher degrees can be obtained by the composition of several
Darboux transformations of degree one.

\section{Darboux transformation in $U(N)$ case}\label{sec3}

When $G=U(N)$, the Lie algebra consists of all anti-Hermitian
matrices. Hence $A_\mu^*=-A_\mu$, $\Phi^*=-\Phi$.

In order to construct Darboux transformation which keeps this
reduction, some constraints on $\zeta_j$'s and $h_j$'s should be
added.

Suppose $\psi$ is a solution of (\ref{eq:lp}), $\phi$ is a
solution of (\ref{eq:lp}) with $\zeta\to\bar\zeta$. Then
\begin{equation} \aligned{l}
   r\partial_r(\phi^*\psi)-2(\zeta-u)\partial_u(\phi^*\psi)=0,\\
   \D 2\partial_v(\phi^*\psi)+\frac{\zeta-u}r\partial_r(\phi^*\psi)=0.
\endaligned \end{equation}
It is uniquely solvable for a given initial value of $\phi^*\psi$ at
$r=r_0>0$ and $v=v_0$. Hence if $\phi^*\psi|_{r=r_0,v=v_0}=0$, then
$\phi^*\psi=0$ identically for $r>0$.

Let $\zeta_0$ be a constant number or a non-constant solution of
(\ref{eq:zetaeq}). Take $Z=\diag(\zeta_1,\cdots,\zeta_N)$ with
$\zeta_j=\zeta_0$ or $\bar\zeta_0$, $H=(h_1,\cdots,h_N)$
where $h_j$ is a column solution of (\ref{eq:lp}) with
$\zeta=\zeta_j$ such that $\det H\ne 0$ and $h_i^*h_j=0$ for
$\zeta_i=\bar\zeta_j$. Then, the Darboux transformation given by
Theorem~\ref{thm1} keeps the $U(N)$ reduction. That is,
$\widetilde A_\mu^*=-\widetilde A_\mu$,
$\widetilde\Phi^*=-\widetilde\Phi$. This is proved similarly as
the $U(N)$ reduction for other systems like the AKNS system.

Darboux transformation of higher degree can be obtained by
composition of Darboux transformations of degree one. However,
when $G=U(N)$, there is the following special and more explicit
construction.

Let $\zeta^{(i)}$ $(i=1,\cdots,r)$ be constant numbers or
non-constant solutions of (\ref{eq:zetaeq}), $h^{(i)}$ be a column
solution of (\ref{eq:lp}) with $\zeta=\zeta^{(i)}$. Consider the
composition of $r$ Darboux transformations of degree one. In the
$i$-th Darboux transformation, let
\begin{equation}
   Z=Z^{(j)}\equiv\diag(\zeta^{(j)},\bar\zeta^{(j)},\cdots,\bar\zeta^{(j)}),
   \quad
   H=H^{(j)}\equiv(h^{(j)}_1,\cdots,h^{(j)}_N)
\end{equation}
where $h^{j}_1=h^{(j)}$ and $h^{(j)}_k$ ($k=2,3,\cdots,n$) are
solutions of (\ref{eq:lp}) with $\zeta=\bar\zeta^{(j)}$ and
satisfy $h_k^{(j)*}h^{(j)}=0$. In this case, the Darboux
transformation of degree $r$ can be constructed in the following
way, which does not depend on $h^{(j)}_k$ $(j=1,2,\cdots,r;\,
k=2,3,\cdots,n)$.

Let
\begin{equation}
   \Gamma_{ij}=\frac{h^{(i)*}h^{(j)}}{\bar\zeta^{(i)}-\zeta^{(j)}},
   \label{eq:Gamma}
\end{equation}
then $G=(G_{ij})$ with
\begin{equation}
   G_{ij}=\prod_{j=1}^r(\zeta-\bar\zeta^{(j)})
   \left(1+\sum_{i,j=1}^r \frac{h^{(i)}(\Gamma^{-1})_{ij}h^{(j)*}}
   {\zeta-\bar\zeta^{(j)}}\right)
   \label{eq:G}
\end{equation}
is a Darboux matrix for (\ref{eq:lp}) \cite{bib:Ward1,bib:Zhou}.

\section{Soliton solutions}\label{sec4}

Soliton solutions are obtained in the following way.

Take seed solution $A_\mu=0$ $(\mu=u,v,r)$, $\Phi=0$.
Considering the gauge equivalence in (\ref{eq:Aueq}), we can
always choose $R=1$ and $T=S$. From
(\ref{eq:Aueq})--(\ref{eq:Phieq}), (\ref{eq:Seq}) and
(\ref{eq:Zeq}), we have
\begin{equation} \aligned{l}
   \D\widetilde A_u=0,\quad
    \widetilde A_v=-(\partial_vS)S^{-1}=\frac 1{2r}\partial_rS,\\
   \D\widetilde A_r=-\frac 12(\partial_rS)S^{-1}=-\frac 1r(\partial_uS+1),\quad
    \widetilde\Phi=-\frac r2(\partial_rS)S^{-1}=-\partial_uS-1,
   \endaligned
   \label{eq:dteg1}
\end{equation}
and
\begin{equation} \aligned{l}
   \D\widetilde F_{uv}=[\widetilde D_u,\widetilde D_v]
    =\frac 1{2r}\partial_u\partial_rS,\\
   \D\widetilde F_{ur}=[\widetilde D_u,\widetilde D_r]
    =-\frac 1r\partial_u\partial_uS,\\
   \D\widetilde F_{vr}=[\widetilde D_v,\widetilde D_r]
    =-\frac 1{2r}(\partial_r\partial_r
    +2\partial_u\partial_v)S+\frac 1{2r^2}\partial_rS
    -\frac 1{2r^2}[\partial_rS,\partial_uS].
   \endaligned \end{equation}

Here we always suppose $Z$ is diagonal with
$Z=\diag(\zeta_1,\cdots,\zeta_N)$. Then the corresponding $h_i$'s
are solved from (\ref{eq:lp}) explicitly.

{\bf Case 1:} $\zeta_i$ is a constant.

Then $h_i$ satisfies
\begin{equation} \aligned{l}
   r\partial_rh_i-2(\zeta_i-u)\partial_uh_i=0,\\
   \D 2\partial_vh_i+\frac{\zeta_i-u}r\partial_rh_i=0.
\endaligned \label{eq:hieq} \end{equation}
Hence
\begin{equation}
   h_i=f_i(\omega(\zeta_i))
\end{equation}
where $f_i$ is an arbitrary holomorphic function of $\omega(\zeta_i)$ and
\begin{equation}
   \omega(\zeta)=v-\frac{r^2}{\zeta-u}.
   \label{eq:omega}
\end{equation}

{\bf Case 2:} $\zeta_i$ is not a constant.

According to (\ref{eq:exprzeta2}), $\zeta_i$ satisfies
\begin{equation}
   v-\frac{r^2}{\zeta_i-u}=C_i(\zeta_i)
\end{equation}
where $C_i$ is an arbitrary holomorphic function. $h_i$ should be
a solution of (\ref{eq:hieq}) with this $\zeta_i$, which is
\begin{equation}
   h_i=f_i(\zeta_i)
\end{equation}
where $f_i$ is an arbitrary holomorphic function. The Darboux
transformation is also given by $S=HZH^{-1}-u$ with
$H=(h_1,\cdots,h_N)$ when $\det H\ne 0$.

Multi-solitons can be obtained by the composition of Darboux
transformations of degree one or by (\ref{eq:Gamma}) and
(\ref{eq:G}) directly in $U(N)$ case.

\section{Examples for SU(2) case}\label{sec5}

Now we consider the soliton solutions for the simplest
non-Abelian group $G=SU(2)$.

{\bf (1) Single soliton solutions with constant spectral parameter}

Take $\zeta_0$ to be a complex constant which is not real,
$Z=\diag(\zeta_0,\bar\zeta_0)$. Let $\tau=\omega(\zeta_0)$, then
\begin{equation}
   H=\pmatrix{cc}\alpha(\tau) &-\overline{\beta(\tau)}\\
   \beta(\tau) &\overline{\alpha(\tau)}\endpmatrix
\end{equation}
where $\alpha$, $\beta$ are two holomorphic functions. Let
$\sigma(\tau)=\beta(\tau)/\alpha(\tau)$, then
\begin{equation}
   S=\frac{\zeta_0-\bar\zeta_0}{1+|\sigma|^2}
   \pmatrix{cc} 1 &\bar\sigma\\ \sigma &|\sigma|^2\endpmatrix+\bar\zeta_0-u.
   \label{eq:egS} \end{equation}
\begin{equation}
   \widetilde\Phi=-\partial_uS-1
    =\frac{\zeta_0-\bar\zeta_0}{(1+|\sigma|^2)^2}
   \pmatrix{cc} (|\sigma|^2)_u &\bar\sigma^2\sigma_u-\bar\sigma_u\\
    \sigma^2\bar\sigma_u-\sigma_u &-(|\sigma|^2)_u\endpmatrix
   \label{eq:egPhi}
\end{equation}
and
\begin{equation}
   -\tr\widetilde\Phi^2=\frac{8(\im \zeta_0)^2}{(1+|\sigma|^2)^2}|
    \partial_u\sigma|^2.
   \label{eq:Energy}
\end{equation}

According to (\ref{eq:hyperboloid}) and (\ref{eq:omega}),
\begin{equation}
   \tau=\frac{\zeta_0(Y+V)(U+X)-1-Y^2+V^2}{(\zeta_0(U+X)-Y+V)(U+X)}
   =\frac{\zeta_0(Y+V)+X-U}{\zeta_0(U+X)-Y+V}.
   \label{eq:tauglobal}
\end{equation}
Denote
\begin{equation}
   \xi=\zeta_0(Y+V)+X-U,\quad
   \eta=\zeta_0(X+U)-Y+V,
\end{equation}
then both $\xi$ and $\eta$ can not be zero anywhere on
(\ref{eq:hyperboloid}) when $\zeta_0$ is not real. Hence $\tau$
is a smooth function of $U,V,X,Y$ on (\ref{eq:hyperboloid}).
Moreover,
\begin{equation}
   \partial_u\tau=-\frac{r^2}{(\zeta_0-u)^2}=-\frac 1{\eta^2}.
\end{equation}
Since $\sigma(\tau)$ is a meromorphic function of $\tau$, suppose
$\sigma(\tau)=\sigma_1(\tau)/\sigma_2(\tau)$ where
$\sigma_1(\tau)$ and $\sigma_2$ are two holomorphic functions of
$\tau$ without common zero. According to (\ref{eq:Energy}),
\begin{equation}
   -\tr\widetilde\Phi^2=
   \frac{8(\im\zeta_0)^2|\sigma_2(\tau)\partial_\tau\sigma_1(\tau)
    -\sigma_1(\tau)\partial_\tau\sigma_2(\tau)|^2}
   {(|\sigma_1(\tau)|^2+|\sigma_2(\tau)|^2)^2}
   |\eta|^{-4}.
\end{equation}
Hence, $\widetilde\Phi$ can be extended smoothly to
(\ref{eq:hyperboloid}). Likewise, according to (\ref{eq:dteg1}),
\begin{equation}
   \aligned{l}
   \D-\tr\widetilde A_u^2=0,\quad
   \D-\tr\widetilde A_v^2=
   \frac{8(\im\zeta_0)^2(|\sigma_2(\tau)\partial_\tau\sigma_1(\tau)
    -\sigma_1(\tau)\partial_\tau\sigma_2(\tau)|^2}
   {(|\sigma_1(\tau)|^2+|\sigma_2(\tau)|^2)^2}
   (U+X)^2|\eta|^{-2},\\
   \D-\tr\widetilde A_r^2=
   \frac{8(\im\zeta_0)^2(|\sigma_2(\tau)\partial_\tau\sigma_1(\tau)
    -\sigma_1(\tau)\partial_\tau\sigma_2(\tau)|^2}
   {(|\sigma_1(\tau)|^2+|\sigma_2(\tau)|^2)^2}
   (U+X)^2|\eta|^{-4}.
   \endaligned
\end{equation}
Therefore, the solution $(\widetilde\Phi,\widetilde
A_u,\widetilde A_v,\widetilde A_r)$ is smooth on
(\ref{eq:hyperboloid}), hence is smooth on the whole \adS.

The infinity of the \adS{} includes only $r\to 0$. However, for
the parameter space $(t,r,x)$ $(r>0)$ with fixed $t$, the
infinity of the derived half plane contains $r\to 0$ and
$r^2+x^2\to \infty$. Here we call a solution localized if
$-\tr\widetilde\Phi^2\to 0$ when $r\to 0$ or $r^2+x^2\to\infty$
for fixed $t$.

\example
If $\zeta_0=\I$, $\tau=\omega(\zeta_0)$, $\sigma(\tau)=\tau$,
this is just the localized solution (25) of \cite{bib:Wardnew},
and
$$ -\tr\widetilde\Phi^2=\frac{8r^4}{((r^2+x^2-t^2)^2+2x^2+2t^2+1)^2}. $$

Let
\begin{equation}
   x=tR\cos\theta,\quad r=tR\sin\theta.
\end{equation}
When $t$ and $\theta$ are fixed, $-\tr\widetilde\Phi^2$ is a
function of $R$ only. Its maximum appears at $R=\pm\sqrt{t^2+1}/t$.
Hence as $t\to\infty$, the ridge of the solution locates on the
circle $x^2+r^2=t^2+1$.

Fig.~\ref{fig:figure1} and Fig.~\ref{fig:figure2} describe this soliton
at $t=0$ and $t=10$ respectively. In these two figures, the
vertical axis is $(-\tr\widetilde\Phi^2)^{1/4}$.

\endexample

\unitlength=1mm

\vbox{%
\ifx\figuretype\BMPfile
{
\begin{picture}(60,50)
\put(6,50){\special{em:graph zhou1.bmp}}
\end{picture}
\vskip0.6cm
}
\else
{
\vskip-1cm
\epsffile[0 0 400 230]{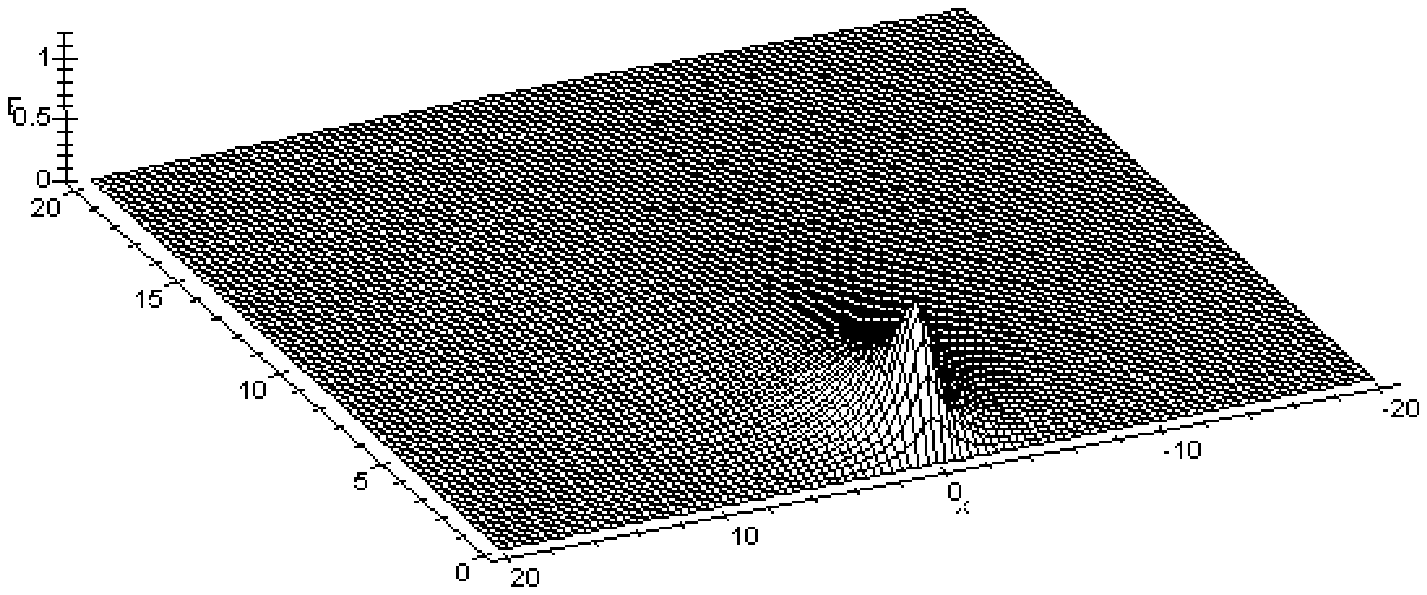}
\vskip-0.5cm
}
\fi
\hbox to \hsize{\hfill\scriptsize Fig.~\ref{fig:figure1}\hfill\hfill}}

\vbox{%
\ifx\figuretype\BMPfile
{
\begin{picture}(60,50)
\put(6,50){\special{em:graph zhou2.bmp}}
\end{picture}
\vskip0.6cm
}
\else
{
\vskip-1cm
\epsffile[0 0 400 222]{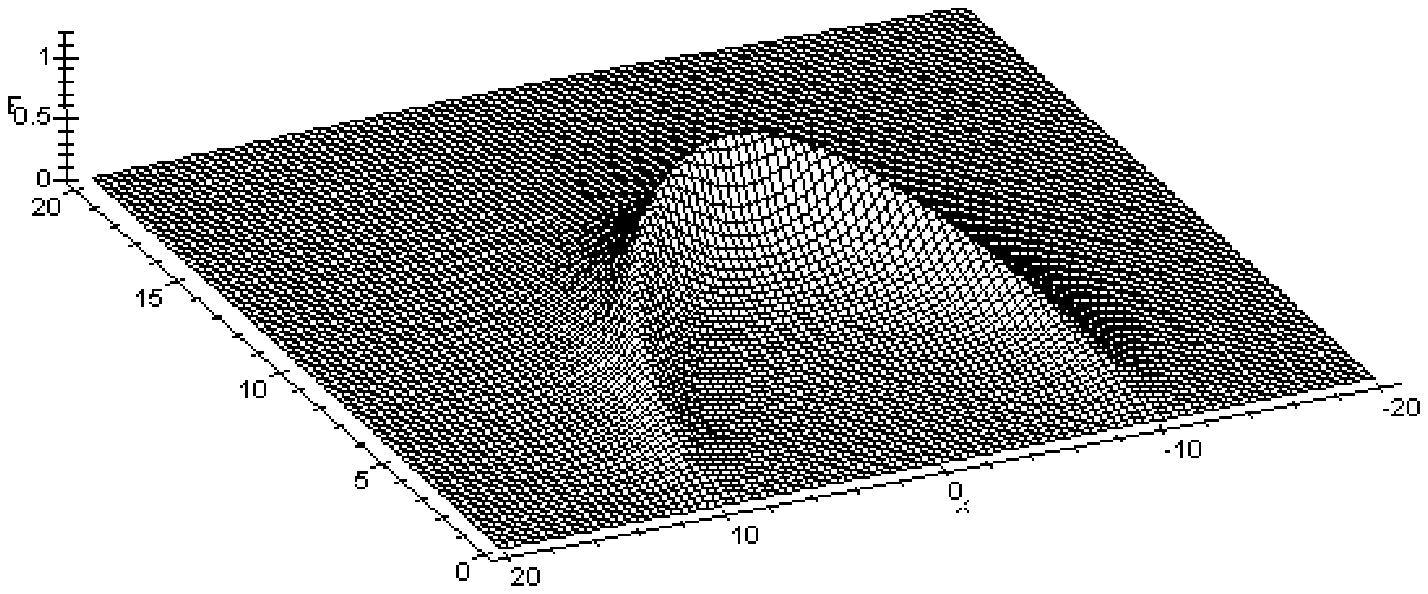}
\vskip-0.5cm
}
\fi
\hbox to \hsize{\hfill\scriptsize Fig.~\ref{fig:figure2}\hfill\hfill}}

\example
$\tau=\omega(\zeta_0)$, $\sigma(\tau)$ is a polynomial of $\tau$ of
degree $k$ $(k\ge 1)$.

If $r\to 0$, then $\tau\to v$, $\partial_v\tau\to 1$ and all the
other derivatives of $\tau$ (including derivatives of higher
orders) tend to zero. According to (\ref{eq:Energy}),
$-\tr\widetilde\Phi^2\to 0$.

If $r^2+x^2\to \infty$, then (for fixed $t$),
\begin{equation}
   \tau=\frac{\zeta_0(x-t)-(r^2+x^2-t^2)}{\zeta_0-x-t}\to\infty.
\end{equation}
Moreover, when $t$ fixed, $r^2+x^2$ and $|\tau|$ are large enough,
we have the estimates
\begin{equation}
   \aligned{l}
   \D\frac{|\tau|^{k+1}|\partial_\tau\sigma|}{1+|\sigma|^2}\le
     C_1\frac{|\tau|^{k+1}|\tau|^{k-1}}{1+|\tau|^{2k}}\le C_1,\\
   \D\left|\frac{\partial_u\tau}{\tau^2}\right|(r^2+x^2)
    =\frac{r^2}{|\zeta_0(x-t)-(r^2+x^2-t^2)|^2}(r^2+x^2)\le C_2
   \endaligned
   \label{eq:esti1}
\end{equation}
where $C_1$ and $C_2$ are independent of $x$ and $r$, but may
depend on $\zeta_0$ and $t$. According to (\ref{eq:Energy}),
\begin{equation}
   -\tr\widetilde\Phi^2\le\frac{8(\im\zeta_0)^2C_1^2C_2^2}{|\tau|^{2k-2}}
   \frac 1{(x^2+r^2)^2}\to 0.
\end{equation}
Hence the solution is also localized whenever $\sigma(\tau)$ is a
non-constant polynomial of $\tau$.

Now we consider the asymptotic behavior of the solution as $t\to
\infty$ for $\zeta_0=\I$. The following discussion in this
example is qualitative and not rigorous.

Suppose all the roots of $\sigma(\tau)$ are simple roots.
Denote $E=-\tr\widetilde\Phi^2$. By (\ref{eq:Energy}), when
$\tau$ is near a root of $\sigma(\tau)$, $E$ may be large. 

From (\ref{eq:omega}), the real and imaginary parts of
$\tau=\omega(\I)$ are
\begin{equation}
   \aligned{l}
   \D\re\tau=\frac{x-t+(x+t)(r^2+x^2-t^2)}{1+(x+t)^2},\\
   \D\im\tau=\frac{r^2}{1+(x+t)^2}.
   \endaligned
   \label{eq:reim_tau}
\end{equation}
When $t$ is large and $x+t$ is not very small,
\begin{equation}
   \re\tau\approx \frac{r^2+x^2-t^2}{x+t}.
\end{equation}
For a root $\rho$ of $\sigma(\cdot)$, the points with
\begin{equation}
   \frac{r^2+x^2-t^2}{x+t}=\re\rho
\end{equation}
are on the circle
\begin{equation}
   C:\quad r^2+(x-\frac 12\re\rho)^2=(t+\frac 12\re\rho)^2.
\end{equation}

On this circle $C$, for fixed $t$ and $\re\rho$, $\im\tau$ can be
expressed by $x$ as
\begin{equation}
   \im\tau=\frac{t^2-x^2+\re\rho(x+t)}{1+(x+t)^2}.
\end{equation}

By computing $\D\frac{\d}{\d x}\im\tau$, we know that $\im\tau$
decreases when $x$ increases if $x\ge -t+1$ and $t\ge-\re\rho/2$.
Hence it is easy to derive that $|\im\tau|\le 2$ when
$t>|\re\rho|$ and $x\ge 0$. Therefore, when $|\im\rho|$ is not
large, there will be a ridge of $E$ on $C$.

When $|\im\rho|>>1$, $E$ is large on $C$ only when
$\im\tau\approx\im\rho$. If $t$ is large and $x+t$ is not very
small,
\begin{equation}
   \im\tau\approx\frac{t-x+\re\rho}{x+t}.
\end{equation}
The equation
\begin{equation}
   \frac{t-x+\re\rho}{x+t}=\im\rho
\end{equation}
has a unique solution
\begin{equation}
   x=\frac{-(\im\rho-1)t+\re\rho}{\im\rho+1}.
\end{equation}
If $\im\rho>>1$, then when $t$ is large enough, $x$ satisfies
$-t\le x\le t+\re\rho$. Hence there exists unique $r>0$ such that
$(x,r)\in C$. This means that when $\im\rho>>1$, there will be a
peak rather than a ridge. If $\im\rho<<-1$, then when $t$ is
large enough, $x<-t$. Hence there does not exist $r>0$ such that
$(x,r)\in C$, that is, there is neither ridge nor peak in the
graph of $E$.

The above discussion on the graph of $E$ is summarized as
follows. As $t\to\infty$, a root $\rho$ with $|\im\rho|<<1$
corresponds to a ridge, a root $\rho$ with $\im\rho>>1$
corresponds to a peak and a root $\rho$ with $\im\rho<<-1$
corresponds to nothing.

Fig.~\ref{fig:figure3} $(t=10)$ shows the solution for
\begin{equation}
   \sigma(\tau)=(\tau-2)(\tau-6)(\tau+6)
\end{equation}
which has three real roots. It is plotted for $r\ge 4$ because the
ridge is perpendicular to $r=0$ and the figures cannot be plotted
well near $r=0$. Fig.~\ref{fig:figure4} shows its local behavior
for a part of the region with $0\le r \le 4$.

\vbox{%
\ifx\figuretype\BMPfile
{
\begin{picture}(60,55)
\put(6,55){\special{em:graph zhou3.bmp}}
\end{picture}
\vskip0.5cm
}
\else
{
\vskip-0.5cm
\epsffile[0 0 400 222]{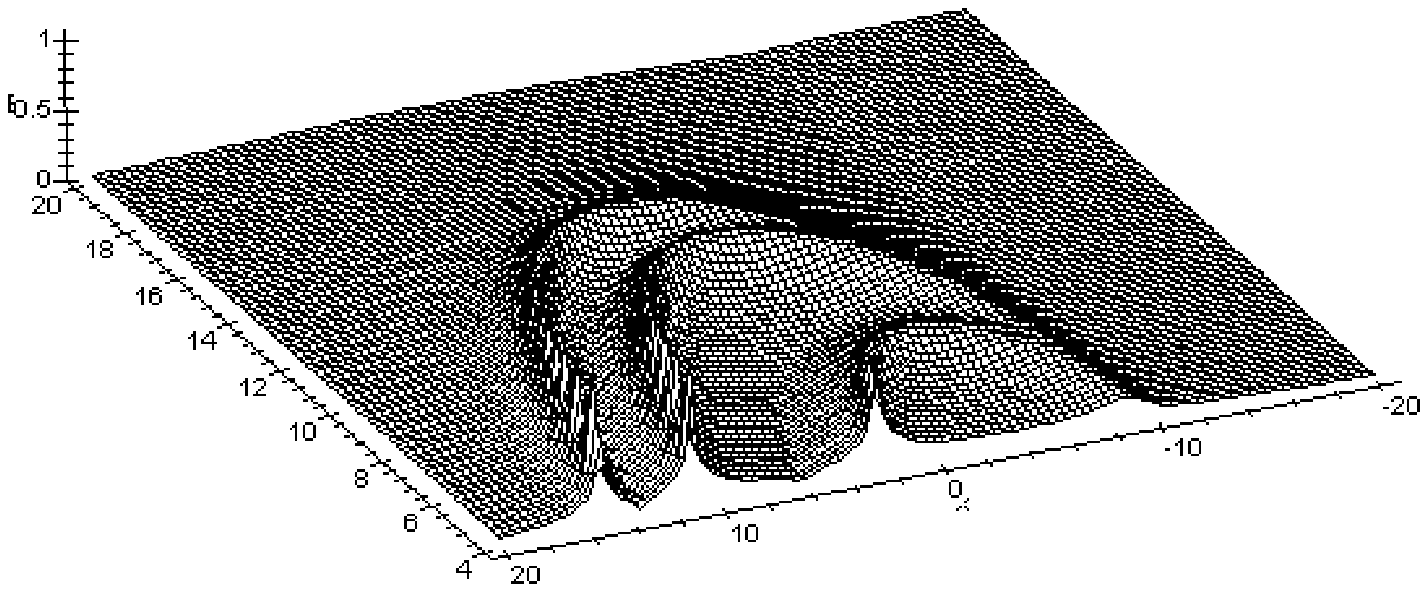}
\vskip-0.5cm
}
\fi
\hbox to \hsize{\hfill\scriptsize Fig.~\ref{fig:figure3}\hfill\hfill}}

\vbox{%
\ifx\figuretype\BMPfile
{
\begin{picture}(60,60)
\put(6,60){\special{em:graph zhou4.bmp}}
\end{picture}
\vskip0.5cm
}
\else
{
\vskip0cm
\epsffile[0 0 400 222]{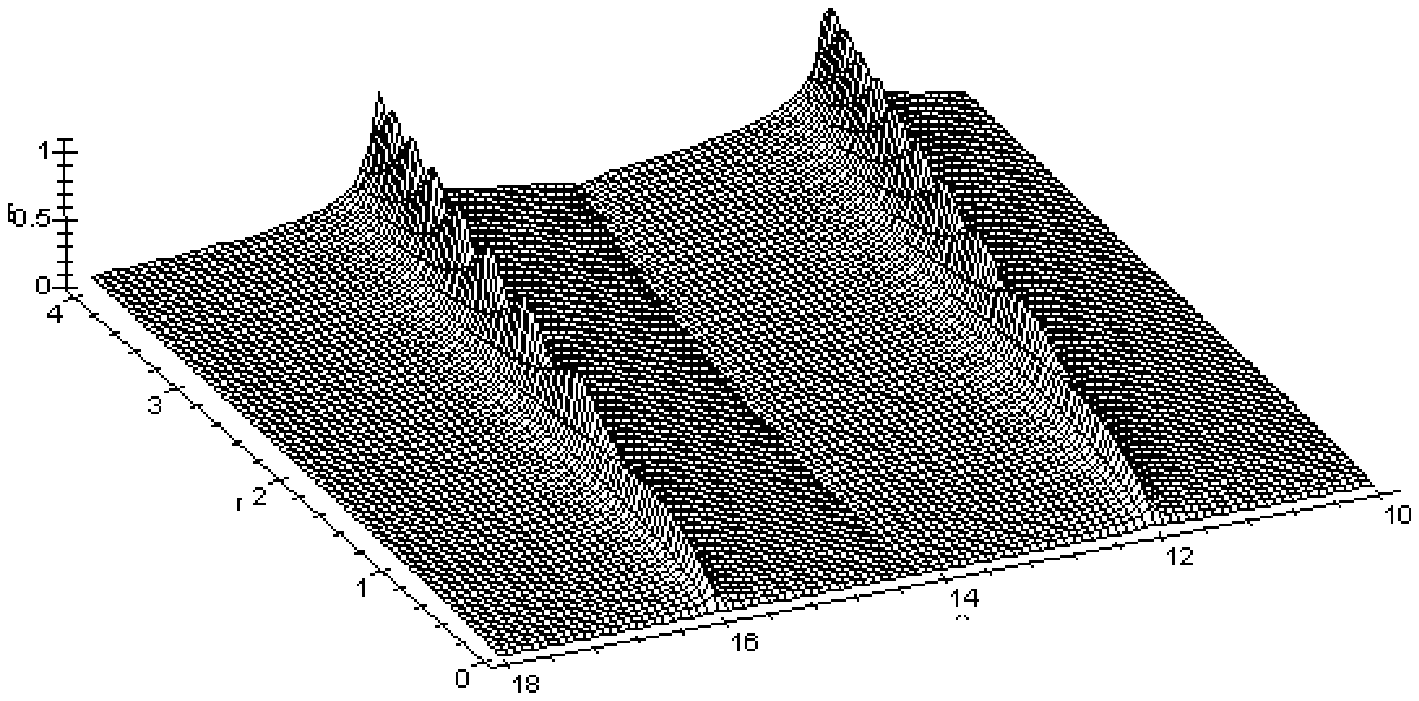}
\vskip-0.5cm
}
\fi
\hbox to \hsize{\hfill\scriptsize Fig.~\ref{fig:figure4}\hfill\hfill}}

Fig.~\ref{fig:figure5} $(t=10)$ shows the solution for
\begin{equation}
   \sigma(\tau)=(\tau-2)(\tau-6)(\tau+6)(\tau-2\I)(\tau-6\I)(\tau+6\I)
\end{equation}
which has three real roots and three purely imaginary roots, but
one imaginary root has negative imaginary part. The solution has
three ridges and two peaks.

\vbox{%
\ifx\figuretype\BMPfile
{
\begin{picture}(60,50)
\put(6,50){\special{em:graph zhou5.bmp}}
\end{picture}
\vskip0.5cm
}
\else
{
\vskip-1cm
\epsffile[0 0 400 222]{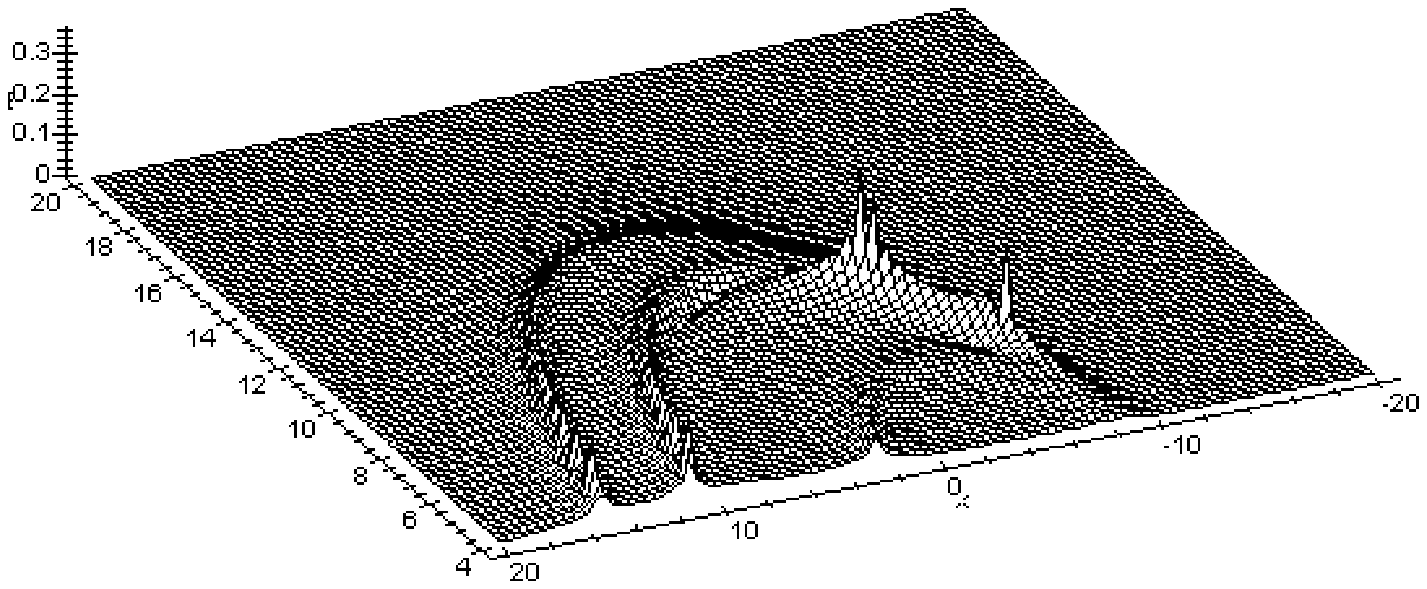}
\vskip-0.5cm
}
\fi
\hbox to \hsize{\hfill\scriptsize Fig.~\ref{fig:figure5}\hfill\hfill}}

Fig.~\ref{fig:figure6} $(t=10)$ shows the solution for
\begin{equation}
   \sigma(\tau)=(\tau-2-2\I)(\tau-6-6\I)(\tau+6-4\I)
\end{equation}
which has no real roots. In the figure, there are three peaks.

\vbox{%
\ifx\figuretype\BMPfile
{
\begin{picture}(60,50)
\put(6,50){\special{em:graph zhou6.bmp}}
\end{picture}
\vskip0.8cm
}
\else
{
\vskip-1cm
\epsffile[0 0 400 222]{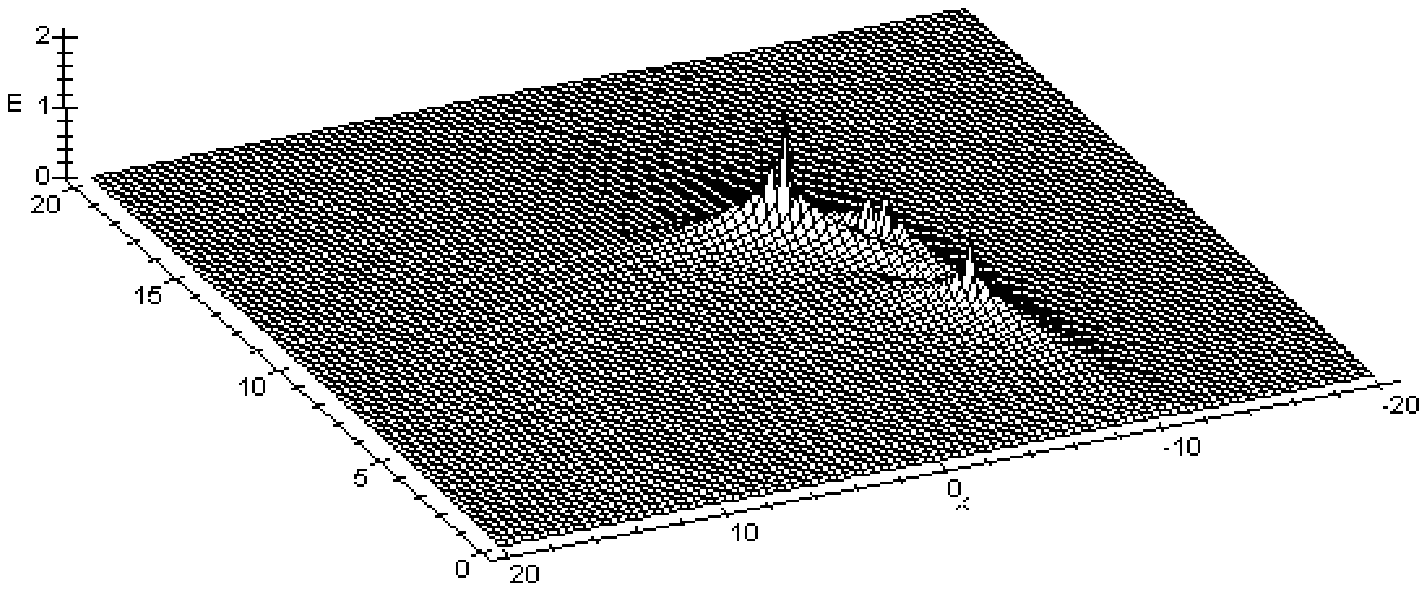}
\vskip-0.5cm
}
\fi
\hbox to \hsize{\hfill\scriptsize Fig.~\ref{fig:figure6}\hfill\hfill}}

Fig.~\ref{fig:figure7} $(t=10)$ shows the solution for
\begin{equation}
   \sigma(\tau)=(\tau-2-2\I)(\tau-6-6\I)(\tau+6+6\I)
\end{equation}
which has no real roots, and only two roots have positive
imaginary parts. In the figure, there are only two peaks.

In all these figures (Fig.~\ref{fig:figure3} to
Fig.~\ref{fig:figure7}), the vertical axis is
$(-\tr\widetilde\Phi^2)^{1/8}$.

\vbox{%
\ifx\figuretype\BMPfile
{
\begin{picture}(60,50)
\put(6,50){\special{em:graph zhou7.bmp}}
\end{picture}
\vskip0.8cm
}
\else
{
\vskip-1cm
\epsffile[0 0 400 222]{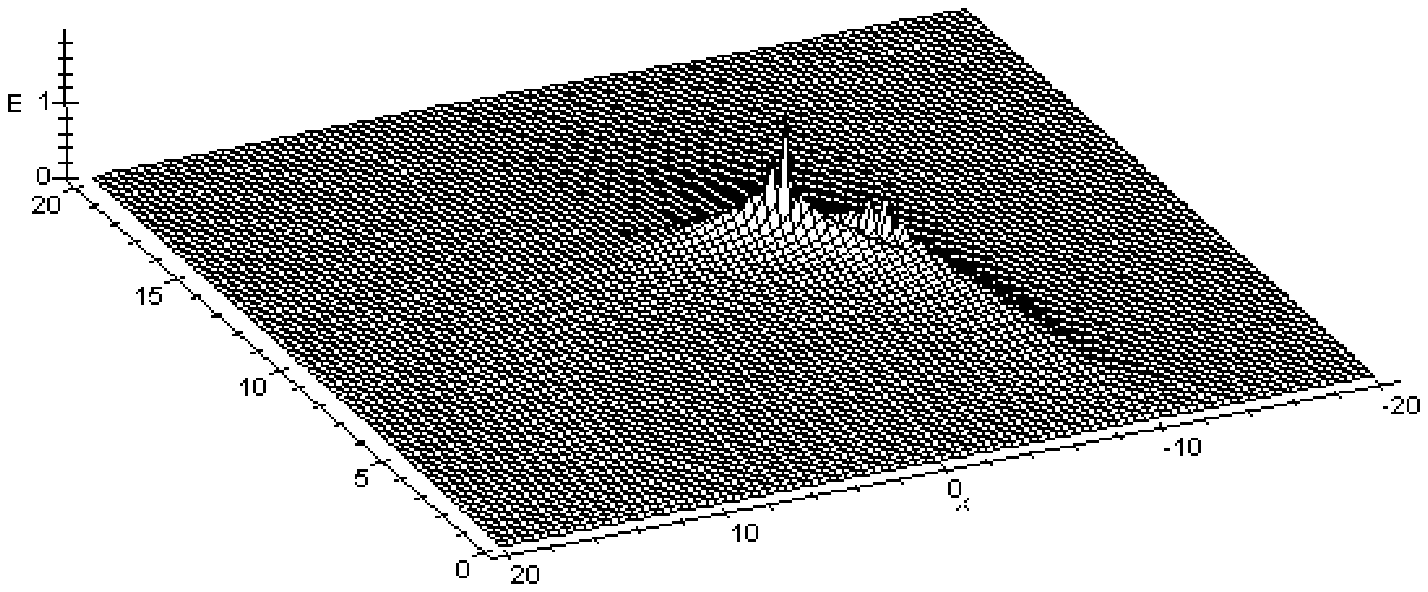}
\vskip-0.5cm
}
\fi
\hbox to \hsize{\hfill\scriptsize Fig.~\ref{fig:figure7}\hfill\hfill}}

\endexample

\example
$\zeta_0=\I$, $\tau=\omega(\I)$, $\sigma(\tau)=\sin(\mu\tau)$
where $\mu$ is a real constant.

If $r\to 0$, then $\tau\to v$, hence $E\to 0$.
When the point $(x,r)\to\infty$ along the straight line $r=kx+b$
($k,b$ are real constants), (\ref{eq:reim_tau}) gives
\begin{equation}
   \aligned{l}
   \D\im\tau=\frac{(kx+b)^2}{1+(x+t)^2}\to k^2,\\
   \D\re\tau=\frac{x-t+(x+t)((kx+b)^2+x^2-t^2)}{1+(x+t)^2}
    \sim(k^2+1)x.
   \endaligned
\end{equation}
Denote $\mu\tau=p+q\I$ where $p$ and $q$ are real, then
\begin{eqnarray}
   E&&=\frac{16\mu^2(\cosh(2q)+\cos(2p))}{(\cosh(2q)-\cos(2p)+2)^2}
    \frac{r^4}{(1+(x+t)^2)^2}\nonumber\\
   &&\sim\frac{16\mu^2k^4(\cosh(2\mu k^2)+\cos(2\mu(k^2+1)x)}
    {(\cosh(2\mu k^2)-\cos(2\mu(k^2+1)x)+2)^2}
\end{eqnarray}
as $x^2+r^2\to\infty$. Hence the solution is bounded, but not
localized in our definition (on the half $(r,x,t)$ space).
However, as is shown, $E$ tends to zero at the infinity of the
\adS{} ($r=0$).

This solution is shown in Fig.~\ref{fig:figure8} ($t=10$).

\vbox{%
\ifx\figuretype\BMPfile
{
\begin{picture}(60,50)
\put(6,50){\special{em:graph zhou8.bmp}}
\end{picture}
\vskip1cm
}
\else
{
\vskip1cm
\epsffile[0 0 400 222]{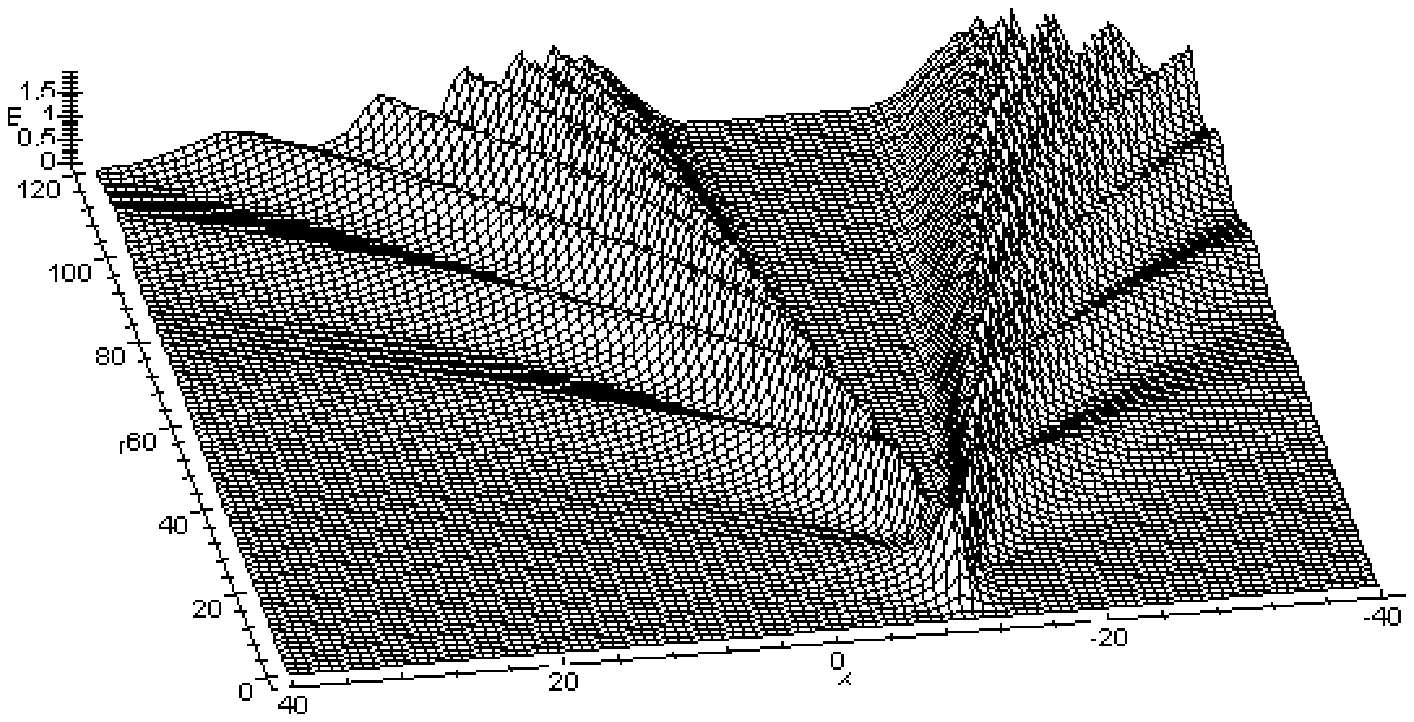}
\vskip-0.5cm
}
\fi
\hbox to \hsize{\hfill\scriptsize Fig.~\ref{fig:figure8}\hfill\hfill}}

\endexample

\vskip6pt
{\bf (2) Single soliton solutions with non-constant ``spectral
parameter''}

In this case, $\zeta_0$ should satisfy
\begin{equation}
   v-\frac{r^2}{\zeta_0-u}=C(\zeta_0).
\end{equation}
$S$, $\widetilde\Phi$ are given as above by (\ref{eq:egS})
and (\ref{eq:egPhi}), and $-\tr\widetilde\Phi^2$ is given by
(\ref{eq:Energy}). However, in these expressions, $\zeta_0$ is no
longer a constant.

Contrary to the case where $\zeta_0$ is a constant, here the
solutions are defined only on the half $(r,x,t)$ space. In
general, they can not be extended to the whole \adS.

\example
$C(\zeta_0)=C_0$ (constant), then
\begin{equation}
   \zeta_0=u-\frac{r^2}{C_0-v}.
\end{equation}

Let
$$ H=\pmatrix{cc}\alpha(\zeta_0) &-\overline{\beta(\zeta_0)}\\
   \beta(\zeta_0) &\overline{\alpha(\zeta_0)}\endpmatrix $$
where $\alpha$ and $\beta$ are holomorphic functions of $\zeta_0$
and $\sigma=\beta(\zeta_0)/\alpha(\zeta_0)$, then the Darboux
matrix $S$ is also given by (\ref{eq:egS}). For example, if
$\sigma(\zeta)=\zeta$ and $C_0=\I$, then $-\tr\widetilde\Phi^2$ is
completely the same as that in Example~1.
\endexample

\example
$C(\zeta_0)=\zeta_0+C_0$ where $C_0=\alpha+\beta\I$, $\alpha$ and
$\beta$ are real constants with $\beta\ne 0$,
$\sigma(\zeta_0)=\zeta_0$.

Then
\begin{equation}
   \zeta_0^2-(u+v-C_0)\zeta_0+uv+r^2-C_0u=0.
\end{equation}
The criteria of this quadratic equation is
\begin{eqnarray}
   \Delta&&=(u+v-C_0)^2-4(uv+r^2-C_0u)
    =(u-v+C_0)^2-4r^2\nonumber\\
   &&=(2t+\alpha)^2-4r^2-\beta^2+2\beta(2t+\alpha)\I.
\end{eqnarray}
When $t>-\alpha/2$, the imaginary part of $\Delta$ is never zero.
Hence we can choose
\begin{equation}
   \zeta_0=\frac{u+v-C_0+\sqrt{(u-v+C_0)^2-4r^2}}2
   \label{eq:snofnonshock}
\end{equation}
where the square root takes the specific branch in the upper plane.

If $r\to 0$, then $\zeta_0\to u$, $\partial_u\zeta_0\to 1$ and
all the other derivatives of $\zeta_0$ tend to zero. Hence by
(\ref{eq:Energy}), $-\tr\widetilde\Phi^2\to 0$.

If $x^2+r^2\to\infty$ ($t$ fixed),
\begin{equation}
   \aligned{l}
   \D\frac{|\zeta_0|}{\sqrt{x^2+r^2}}
   =\frac 1{\sqrt{x^2+r^2}}\left|
   \frac{2x-C_0+\sqrt{(2t+C_0)^2-4r^2}}2\right|\to 1,\\
   \D|\partial_u\zeta_0|=\frac 12\left|
   1+\frac{2t+C_0}{\sqrt{(2t+C_0)^2-4r^2}}
   \right|\le C_3,
   \quad \frac{|\im\zeta_0|}{r}\le 2
   \endaligned
   \label{eq:estimatevar}
\end{equation}
where $C_3$ is independent of $x$ and $r$, but may depend on $t$,
$\alpha$ and $\beta$. According to (\ref{eq:Energy}) for
$\sigma(\zeta_0)=\zeta_0$, $-\tr\widetilde\Phi^2\to 0$ as
$x^2+r^2\to\infty$. Hence the solution is also localized.
However, it can not be extended to the whole \adS{} smoothly
because of the condition $t>-\alpha/2$.

This soliton is shown in Fig.~\ref{fig:figure9} ($t=1$) and
Fig.~\ref{fig:figure10} ($t=10$) for $\alpha=0$ and $\beta=2$.

\vbox{%
\ifx\figuretype\BMPfile
{
\begin{picture}(60,50)
\put(6,50){\special{em:graph zhou9.bmp}}
\end{picture}
\vskip0.5cm
}
\else
{
%\vskip-1cm
\epsffile[0 0 400 222]{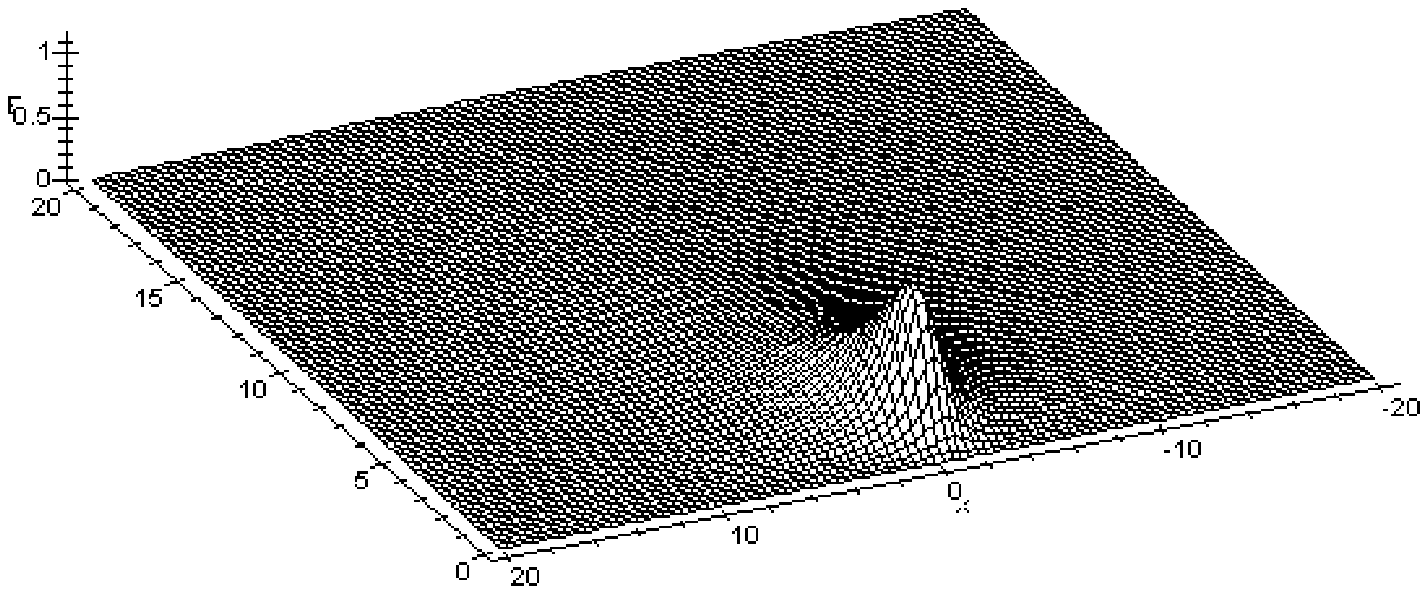}
\vskip-0.5cm
}
\fi
\hbox to \hsize{\hfill\scriptsize Fig.~\ref{fig:figure9}\hfill\hfill}}

\vbox{%
\ifx\figuretype\BMPfile
{
\begin{picture}(60,50)
\put(6,50){\special{em:graph zhou10.bmp}}
\end{picture}
\vskip0.5cm
}
\else
{
\vskip-1cm
\epsffile[0 0 400 222]{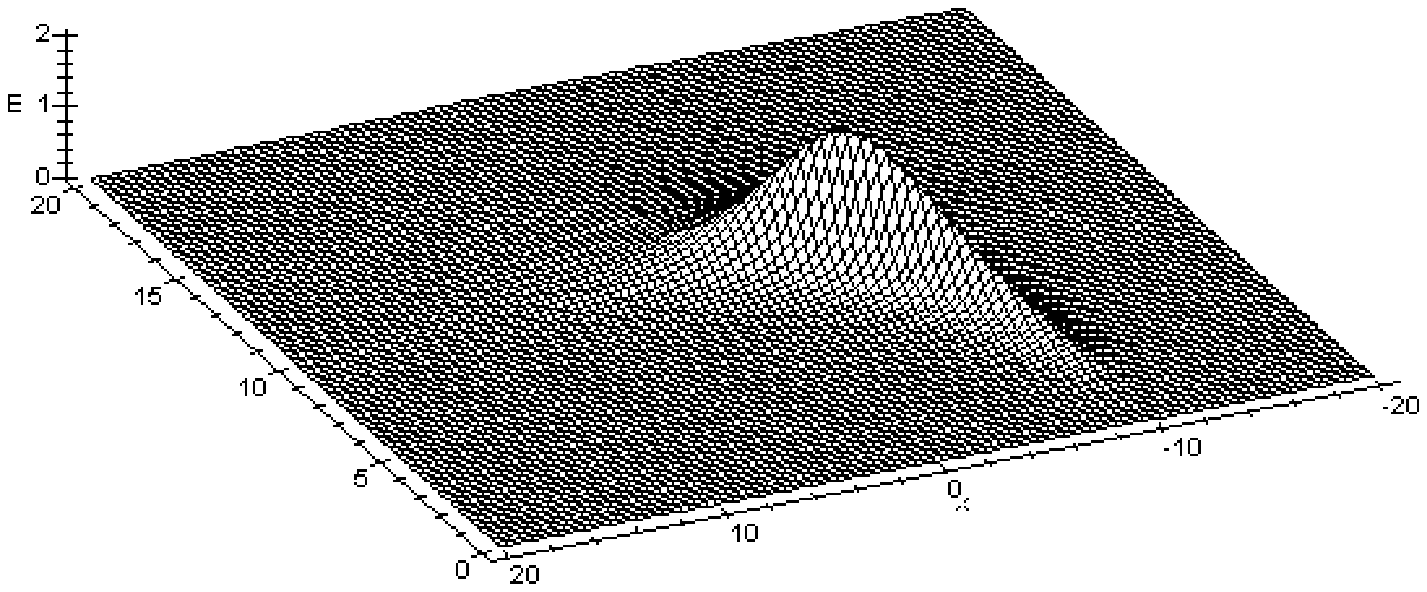}
\vskip-0.5cm
}
\fi
\hbox to \hsize{\hfill\scriptsize Fig.~\ref{fig:figure10}\hfill\hfill}}

\endexample

\example
$C(\zeta_0)=\zeta_0+C_0$, $C_0=\alpha+\beta\I$ $(\beta\ne 0)$ as
above, $\sigma(\zeta_0)$ is a polynomial of $\zeta_0$ of degree
$k$.

Then similar to (\ref{eq:esti1}), we have the estimates
\begin{equation}
   \aligned{l}
   \D\frac{|\zeta_0|^{k+1}|\partial_{\zeta_0}\sigma|}{1+|\sigma|^2}\le
     C_1\frac{|\zeta_0|^{k+1}|\zeta_0|^{k-1}}{1+|\zeta_0|^{2k}}\le C_1,\\
   \D\left|\frac{\partial_u\zeta_0}{\zeta_0^2}\right|(r^2+x^2)
    \le 2C_3
   \endaligned
   \label{eq:esti2}
\end{equation}
as $x^2+r^2\to \infty$ by (\ref{eq:estimatevar}). From
(\ref{eq:Energy}), when $x^2+r^2\to\infty$,
\begin{equation}
   -\tr\widetilde\Phi^2\le \frac{128C_1^2C_3^2}{|\zeta_0|^{2k-2}}
   \frac{r^2}{(r^2+x^2)^2}\to 0.
\end{equation}
The solution is localized. As in the last example, this solution can
not be extended to the whole \adS.

\endexample

{\bf (3) Double soliton solutions}

Double soliton solutions are obtained by Darboux
transformations of degree two. Here we only show the following
simple case for constant spectral parameters. In more complicated
cases the double soliton solutions can also be derived similarly.

\example
Let $\zeta_1=\I$, $\zeta_2=5+2\I$, $\tau_j=\omega(\zeta_j)$,
$\sigma_j(\tau_j)=\tau_j$ $(j=1,2)$. The double soliton is shown in
Fig.~\ref{fig:figure11} ($t=10$).

\vbox{%
\ifx\figuretype\BMPfile
{
\begin{picture}(60,50)
\put(6,50){\special{em:graph zhou11.bmp}}
\end{picture}
\vskip1cm
}
\else
{
\vskip0cm
\epsffile[0 0 400 222]{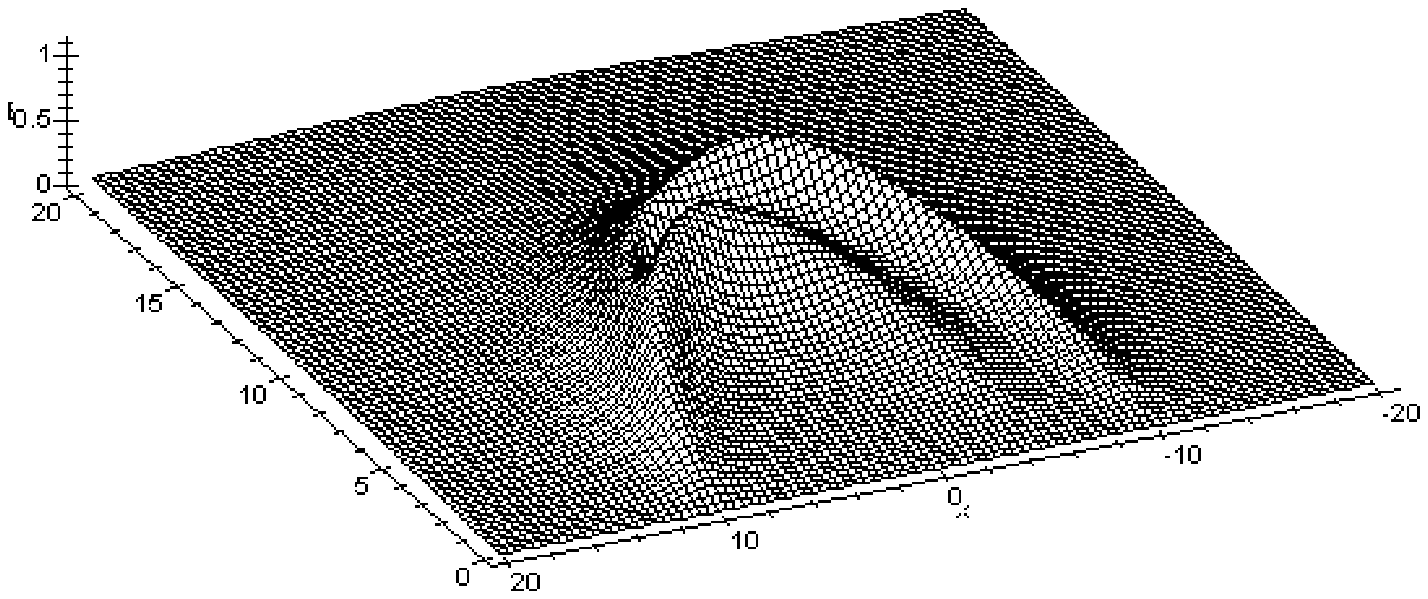}
\vskip-0.5cm
}
\fi
\hbox to \hsize{\hfill\scriptsize Fig.~\ref{fig:figure11}\hfill\hfill}}
\endexample

\section*{Acknowledgements}
This work was supported by Chinese National Research Project
``Nonlinear Science'', the Doctoral Program Foundation and the
Foundation for University Key Teacher by the Ministry of
Education of China. The author is very grateful to
Prof.~M.~F.~Atiyah and Prof.~M.~L.~Ge for the suggestions
and discussions. He would also like to thank the referee
for valuable comments.

\newpage

\begin{figure}
\caption{}
\label{fig:figure1}
\end{figure}

\begin{figure}
\caption{}
\label{fig:figure2}
\end{figure}

\begin{figure}
\caption{}
\label{fig:figure3}
\end{figure}

\begin{figure}
\caption{}
\label{fig:figure4}
\end{figure}

\begin{figure}
\caption{}
\label{fig:figure5}
\end{figure}

\begin{figure}
\caption{}
\label{fig:figure6}
\end{figure}

\begin{figure}
\caption{}
\label{fig:figure7}
\end{figure}

\begin{figure}
\caption{}
\label{fig:figure8}
\end{figure}

\begin{figure}
\caption{}
\label{fig:figure9}
\end{figure}

\begin{figure}
\caption{}
\label{fig:figure10}
\end{figure}

\begin{figure}
\caption{}
\label{fig:figure11}
\end{figure}
\end{document}